\hsize=15.5 truecm
\vsize=22.5 truecm
\leftskip=1 truecm
\topskip=1 truecm
\splittopskip=1 truecm
\parskip=0 pt plus 1 pt
\baselineskip=17.9 pt
\def\re{$\rm I \kern-0.07cm R$}

\def\o{\omega} 
\def\a{\alpha}
\def\b{\beta}
\def\d{\partial}
\def\r{\rho}
 
\vskip 3 cm
\centerline{\bf GENERATING NEW PERFECT-FLUID SOLUTIONS }
\centerline{\bf FROM KNOWN ONES}
\bigskip
\centerline{{\sl Istv\'an R\'acz}
{\footnote{$^{\dag}$}{\sevenrm E-mail: 
istvan@rmkthe.rmki.kfki.hu}} 
and {\sl J\'ozsef Zsigrai}
{\footnote{$^{\ddag}$}{\sevenrm E-mail:
zsigrai@hal9000.elte.hu}}}

\centerline{MTA KFKI Research Institute for Particle and
Nuclear Physics}
 
\centerline{H-1525 Budapest 114, P.O.B. 49, Hungary }

\bigskip
\bigskip
\centerline{\bf \sl Abstract }

{\it Stationary perfect-fluid configurations of Einstein's 
theory of gravity are studied.
It is assumed that the 4-velocity of the fluid is parallel to
the stationary Killing field, and also that the norm and the
twist potential of the stationary Killing field are functionally
independent. It has been pointed out earlier by one of us (I.R.)
that for these perfect-fluid geometries some of
the basic field equations are invariant under an $SL(2,$\re$)$
transformation. Here it is shown 
that this transformation can be applied to generate possibly new perfect-fluid
 solutions from existing known ones only for the case of 
barotropic equation of state $\rho+3p=0$.
In order to study the effect of 
this transformation, its application to known perfect-fluid
 solutions is presented. In this way, different previously 
known solutions could be written in a singe compact form.
A new derivation of all Petrov type D 
stationary axisymmetric rigidly rotating perfect-fluid solutions 
with an equation of state $\rho+3p=constant$ is given in an appendix.}

\bigskip
\parindent 0 pt
PACS number: 04.20.Jb, 04.40.+c
\bigskip

\hsize 15.5 truecm
\leftskip 1 truecm
\bigskip

{\bf 1. Introduction}
\parindent 20 pt
\medskip
 
In spite of great efforts, no stationary axisymmetric perfect-fluid solution of 
Einstein's equations has been
found which would be appropriate for
describing, in the framework of general relativity, rapidly-rotating
 compact massive stars. In fact, only a small number of stationary axisymmetric 
perfect-fluid solutions without higher symmetries are known 
[1-8] (see also references therein). On the other hand,
owing to the nonlinear character of 
Einstein's equations, only the knowledge of as much solutions as 
possible could provide us with a deeper insight into
the methods that might help in 
obtaining astrophysically relevant solutions of
Einstein's 
equations. Therefore, every new solution generating
method is
welcome in general relativity.

Here we would like to present and apply a technique by which one can generate
new solutions from previously known ones.
This technique is, in fact, a generalization of a method
given by Geroch [9] by which out of any
source-free solution of Einstein's equations possessing a non-null 
Killing vector
field one could get a new
one-parameter family of vacuum solutions. 
It has been recently shown by one of us [10], that an analogous 
transformation can be successfully used 
for spacetimes possessing a non-null Killing field
and certain kinds of matter fields.
In particular,
some of the basic field equations for perfect-fluid
 matter sources having 4-velocity 
parallel to a timelike Killing field possess exactly the same type 
of invariance as the vacuum field equations [10,11]
\footnote{$^1$}{\sevenrm The invariance properties of the
basic field equations for electrically charged rigid perfect fluids were 
studied earlier by Kramer {\it \sevenrm \it et al} [12, 13]. Note, however, that their considerations were 
restricted to the {\sevenrm \it static} case exclusively. An application 
of their 
approach was the derivation of a regular static charged interior
Reissner-Nordstr\"om solution from the interior Schwarzschild
metric [14]. As far as we know, the results used in the present paper to
analyze the invariance properties of the field 
equations are new and they were published first in 
Refs. [10,11].}.
A perfect fluid with 4-velocity parallel to a
timelike Killing
field is ``rigid", i.e., it is expansion- and shear-free. Consequently, 
such a fluid 
seems to be far from being appropriate for astrophysical applications. 
However, as it was shown by Geroch and Lindblom [15],
perfect fluids of this kind represent equilibrium configurations 
of dissipative relativistic fluids. Thereby, the study
 of these models
is of obvious astrophysical interest.  Note that this ``rigidity" 
assumption
introduces considerable simplifications of the basic field
equations.

In this paper we consider stationary perfect-fluid configurations 
with 4-velocity parallel to the stationary Killing field. 
The norm and the twist potential of the Killing
field are assumed to be functionally independent.
In section 2 we recall some of the basic
notions and results of the projection (or 3-dimensional) formalism
of general relativity developed for spacetimes with a non-null 
Killing field.
Then, we present the field equations in geometrically 
preferred local coordinates.
In section 3 
a transformation by which one can generate new solutions from 
known ones will be recalled. In particular, we determine the conditions
under which this transformation can be applied to the selected perfect-fluid
spacetimes. It turns out that the original perfect fluid
has to possess the barotropic equation of state $\rho+3p=0$
which is invariant under
the action of the transformation.
Finally, in section 4,
we apply the developed technique to all existing suitable
stationary perfect-fluid solutions.
Accordingly, in section 4, first we recall the most general family of
known stationary axisymmetric perfect-fluid solutions for which the
equation of state is $\rho+3P=0$, the flow is rigid in the above sense and
the twist potential and the norm of the Killing field are functionally
independent. These solutions belong to the large family consisting of all
Petrov type D stationary axisymmetric rigidly-rotating perfect-fluid
metrics with the equation of state $\rho+3P=constant$, obtained by
Senovilla [1]. Note that this family contains as a subfamily
each previously known stationary axisymmetric perfect-fluid
geometry with the equation of state $\rho
+3p=constant$.
Several other solutions were found later by Mars and Senovilla which
are also of Petrov 
type D stationary perfect-fluid solutions -- one of 
them possesses two commuting timelike Killing fields while the 
other is a stationary axisymmetric solution -- and
the equation of state is $\rho+3p=0$ for both of them. 
In appendix 1 these solutions are shown to be subclasses of the ones 
given by Senovilla. Finally, a new form of all Petrov 
type D stationary axisymmetric rigidly rotating metrics with 
an equation of state $\rho+3p=constant$ is presented in appendix 2.
The solutions yielded by the application of the generalized 
Geroch transformation to the original Senovilla metrics can be 
related easily to these general but relatively compact and 
simple metrics. Moreover, this family of solutions contains on 
equal footing both of the original classes of the Senovilla 
solutions.

\bigskip
\parindent 0 pt
{\bf 2. Basic notions and the field equations}
\parindent 20 pt
\medskip

We consider perfect-fluid spacetimes, $(M, g_{ab})$, admitting a timelike 
Killing field, $\xi^a$. The mass density, the pressure and the 4-velocity 
of the fluid are denoted by
$\rho$, $p$ and $u^a$, respectively. We assume, furthermore, that 
the 4-velocity
is parallel to the timelike Killing field $\xi^a$, which implies
$$u^a=(-v)^{-{1\over 2}}\xi^a .			\eqno(2.1)$$
Then, the flow is rigid in the sense that it is expansion- and 
shear-free. As mentioned above, this
assumption on rigidity is compatible with the 
general strategy that one would like to find a faithful 
description of equilibrium states of dissipative relativistic fluids.

It is known that the entire geometrical content of a 
spacetime possessing a timelike Killing field can be represented on the 
3-dimensional space of Killing orbits by a triple ($v$, $\o_a$, 
$h_{ab}$)  where $v$ is the norm of the Killing field,
$$v=\xi^a\xi_a , 		\eqno(2.2)$$
$\o_a$ is the twist of the Killing field,
$$\o_a=\epsilon_{abcd}\xi^b\nabla^c\xi^d ,	\eqno(2.3)$$
(here $\epsilon_{abcd}$ is the 4-volume element), while
$h_{ab}$ is a Riemannian metric defined as
$$h_{ab}=-v g_{ab}+\xi_a\xi_b . \eqno(2.4)$$

For the selected stationary perfect-fluid configurations
$\o_a$ can be expressed (at least locally) as the gradient of
a function, $\o$, called the twist potential
$$\o_a=D_a\o .				\eqno(2.5)$$

Throughout this paper we shall assume that the above defined functions 
$v$ and $\o$ are functionally independent, i.e.,
$$D_{[a}\o D_{b]}v\not= 0 .			\eqno(2.6)$$
(The complementary case, when $v$ and $\o$ are functionally dependent, 
will be considered
in a subsequent paper.) Whenever eq. (2.6) is satisfied,
the basic field equations can be 
simplified by the introduction of geometrically preferred local coordinates, 
$(x^1,x^2,x^3)$ [10,11]. Then we have that
the functions $v$ and $\o$ and also --
by virtue of the field equations and the equation of state which is of the form
$\r=\r(p)$ for the considered rigid perfect fluids -- the
functions $\r$ and $p$ depend only on the coordinates 
$x^1$ and $x^2$. In these local coordinates the set of
equations 
describing a stationary perfect fluid with 4-velocity 
parallel to the stationary Killing field reduces to [10,11]
$$2(R_{AB}-{R_{33}\over h_{33}}h_{AB})=
v^{-2}(\d_A v\d_B v+\d_A\o\d_B\o) ,		\eqno(2.7)$$
$$R_{3\b}=16\pi v^{-1}p h_{3\b}	,		\eqno(2.8)$$
$$\d_Ap+{1\over 2} (\r+p){{\d_Av}\over v}=0 ,		
\eqno(2.9)$$
where $R_{ab}$ is the Ricci tensor associated with the three 
dimensional metric $h_{ab}$ and the capital Latin indices take the values $1,2$ while the Greek
ones take $1,2,3$. As it follows from the results
of Refs. [10,11], 
for the stationary perfect-fluid geometries under consideration it is 
sufficient to solve the above (reduced) set of field equations, 
because the functional independence of $v$ and $\o$ ensures that 
any solution of this set satisfies the complete system of Einstein's 
equations.

\bigskip
\parindent 0 pt
{\bf 3. The transformation}
\parindent 20 pt
\medskip

In this section first the symmetry properties of
eq. (2.7) will be studied. In particular, it will be recalled that
a one-parameter family of new solutions of eq. (2.7) can
be associated with any given solution of this equation.
Then, the restrictions on the applicability of the related 
transformation -- imposed by the remaining field equations,
(2.8) and (2.9) -- will be determined.

To start off, note that, just like for the vacuum case,
the left hand side of eq. (2.7) depends merely on the
3-dimensional metric, $h_{ab}$, while the right-hand side
is given in terms of the functions $v$ and $\o$.
By exactly the same argument as used for the vacuum case 
in Ref. [9], it can be shown that by starting with a 
particular solution,
($v_0,\o_0$), of eq. (2.7) associated with a {\it fixed} 
3-dimensional metric, $h_{ab}$, (and thereby, a fixed set 
of functions $R_{AB}-{R_{33}\over h_{33}}h_{AB}$) one can 
generate a one-parameter family of solutions, 
($v_\tau,\o_\tau$), of eq. (2.7). More precisely, 
let ($v_0,\o_0$) be a particular solution; then, the full set of solutions of 
eq. (2.7) associated with a fixed 3-metric, $h_{ab}$,
can be given as
$$v_\tau={v_0\over{(\cos\tau-\o_0 \sin\tau)^2+v_0^2\sin^2\tau}} ,
\eqno(3.1)$$

$$\o_\tau={{(\o_0\cos\tau+\sin\tau)(-\o_0\sin\tau+\cos\tau)-
v_0^2\sin\tau\cos\tau}\over{(\cos\tau-\o_0\sin\tau)^2+v_0^2\sin^2\tau}} .
\eqno(3.2)$$

These formulas, with $\tau$ as the only independent parameter, can 
be derived from the general form of an $SL(2,$\re$)$ transformation by 
factoring out with respect to the trivial two-parameter gauge 
transformations [9].
Although this transformation may yield gauge
inequivalent geometries, which can be shown to form a circle, its repeated
applications do not generate new 
solutions since the transformation then merely yields a rotation 
of this circle \footnote{$^2$}{\sevenrm The properties of the 
transformation are 
analyzed in detail in Ref. [9] for the vacuum case. Exactly the 
same analysis applies for the present case.}.

The most significant difference between the set of vacuum 
field equations and the field equations for the case
under consideration is that for the vacuum problem the appropriate 
form of eq. (2.7) is the only equation to be solved, while for the
perfect-fluid case the basic field variables,
beside eq. (2.7), have to satisfy eqs. (2.8) and (2.9),
as well. These equations, however, impose non-trivial restrictions on the 
applicability of the transformation.

To determine these restrictions, let us consider first eq. (2.8).
It can be easily seen that if one fixes the 3-geometry, $h_{ab}$, 
the quantity $pv^{-1}$ has to be left intact by the 
above transformation. 
While $v$ transforms according to eq. (3.1), $p$ has to transform 
so that the variation of $p$ compensates the variation of $v$.
This implies that 
$${p_\tau \over v_\tau}={p_0\over v_0}	,	\eqno(3.3)$$
has to hold, where $p_0$ and $v_0$ denote the pressure and norm of the Killing
field for the original perfect-fluid configuration,
while $p_\tau$ and $v_\tau$ are the corresponding functions 
for the transformed geometries. This equation represents one 
of the subsidiary
conditions to be satisfied whenever one applies  
the above transformation.

A further restriction
is risen by the Euler-Lagrange equation (2.9). In fact, 
eqs. (3.1) and (3.3) can be used to determine the transformed 
mass density, $\rho_\tau$, and, thereby, the transformation law 
for the equation of state of the fluid. To see this, 
take the derivative of eq. (3.3)
$$v_\tau^{-1}(\d_Ap_\tau-p_\tau{{\d_Av_\tau}\over v_\tau})=
v_0^{-1}(\d_Ap_0-p_0{{\d_Av_0}\over v_0}).	\eqno(3.4) $$
Then, using the Euler-Lagrange equation (2.9),
one gets from (3.4)
$$(\r_\tau+3p_\tau){{\d_Av_\tau}\over v_\tau^2}=
(\r_0+3p_0){{\d_Av_0}\over v_0^2} .	\eqno(3.5)$$

Finally, substituting the right hand side
of (3.1) for $v_\tau$ into (3.5) one obtains,
$$(\r_0+3p_0)\d_Av_0=$$
$$=(\r_\tau+3p_\tau)
\Bigl\{\bigl[(\cos\tau-\o_0 \sin\tau)^2 -\sin\tau^2 v_0^2\bigr] \d_Av_0 
+2 \sin\tau v_0(\cos\tau-\o_0 \sin\tau)\d_A\o_0\Bigr\}	.
\eqno(3.6)$$

Since $v_0$ and $\o_0$ are supposed to be functionally independent, the
coefficients of $\d_Av_0$ and $\d_A\o_0$ in eq. (3.6) must vanish separately. This,
however, implies that both of the quantities $\rho_0+3p_0$ and
$\rho_\tau+3p_\tau$ must vanish identically.
Correspondingly, the above transformation can be applied only to those
perfect-fluid configurations that have the barotropic equation of state 
$\rho+3p=0$.
Furthermore, each of the new solutions yielded by the transformation
has to possess this equation of state.

It is important to emphasize that there is no further restriction on 
the applicability of the above transformation (see eqs. (3.1) and 
(3.2)), risen by eqs. (2.8) and (2.9).
In fact, for these perfect-fluid geometries with
equation of state $\r+3p=0$ the Euler-Lagrange equation can be
integrated. It is easy to check that the general solution of eq. (2.9)
for this equation of state is $pv^{-1}=constant$ which, by virtue of eqs. 
(2.8) and
(3.3), implies (in accordance with the fact that $h_{ab}$ is kept 
fixed) that $p_\tau v_\tau ^{-1}=p_0 v_0^{-1}=constant$.
Hence, for these
perfect-fluid configurations the functional
forms of $\r_\tau$ and $p_\tau$ will not be explicitly presented.
They can be obtained by simply multiplying
$v_\tau$ by constant factors.

\bigskip
\parindent 0pt
{\bf 4. Getting solutions from existing ones}
\parindent 20pt
\medskip

In this section the above transformation is applied 
to known stationary perfect-fluid solutions 
of Einstein's equations with equation of state $\r+3p=0$. The 
main steps of the procedure of getting new solutions
are the following:
Start with a stationary perfect-fluid solution with 
equation of state $\r_0+3p_0=0$. Determine, first, the functions 
$v_0$ and $\o_0$ as well as the fixed 3-metric, $h_{ab}$, which are
the input data for our procedure. Then, 
determine the functional form of $v_\tau$ and $\o_\tau$ 
using eqs. (3.1) and (3.2). Finally, the 4-dimensional line element 
can be given, in virtue of eq. (16.22) of Ref. [13], by the integration
of the relevant form of eq. (16.23) of Ref. [13].

The stationary perfect-fluid solutions which can be used as input for the 
above procedure are, in fact, rare. There is a large class containing  
Petrov type D stationary axisymmetric perfect-fluid solutions with the 
equation of state $\r+3p=constant$, given by Senovilla [1], which includes as a 
special subcase the Wahlquist solution [13] and certain families of 
solutions given earlier by Kramer [5,6] and by Mars and Senovilla [3]
\footnote{$^3$}{\sevenrm It was not mentioned originally by the authors in 
Ref. [3] 
that two of the solutions given there represent, actually, a subclass 
of the previously published Senovilla metrics.}.

The general form of all Petrov type D stationary axisymmetric rigidly 
rotating perfect-fluid metrics with an equation of state 
$\rho+3p=constant$ can be written [1] as
$$ds^2=v_0(dt+A_0 
d\phi)^2+(V-W)[G^{-1}dx^{2}+H^{-1}dy^{2}+c^2GH(G-H)^{-1}d \phi ^{2}]
\eqno(4.1)$$
where the norm of the stationary Killing field, $\bigl({\d\over{\d 
t}}\bigr)^a$,
is 
$$v_0={{H-G}\over{V-W}}	,			\eqno(4.2)$$
furthermore,
$$A_0={{c(HV-GW)}\over{H-G}},			\eqno(4.3)$$
and $c$ is an arbitrary constant.
Here the functions $V=V(x)$, $W=W(y)$, $G=G(x)$ and $H=H(y)$ satisfy eqs. 
(16) and (17) of Ref. [1] (see also eqs. (A2.1) and (A2.2) of 
the present paper). The Senovilla metrics are divided into two 
classes. The first one, class I with ${dV\over{dx}}\not =0$,
coincides for a special setting of the parameters [1]  
with the Wahlquist solution [13], while in 
class II with ${dV\over{dx}}=0$, one can also find metrics given  previously by Kramer
[5,6] and by Mars and Senovilla [3].  For the Senovilla metrics 
the functions $V$, 
$W$, $G$ and $H$ satisfy eqs. (18), (19a) and (19b) of Ref. [1] for class I 
and eqs. 
(18), (20a) and (20b) of Ref. [1] for class II. 
The general solutions of 
these equations for the case of $\rho+3p=0$ 
(which corresponds to the vanishing of the parameter $a$ in these equations) 
can be given as follows:

For class I:
$$V(x)=m_0+M \sin(4bx+x_0) \eqno(4.4)$$
$$W(y)=m_0+M\cosh(4by+y_0) \eqno(4.5)$$
$$G(x)=s_0+S\sin(4bx+x_0)+N_1\cos(4bx+x_0)	\eqno(4.6)$$
$$H(y)=s_0+N_2\sinh(4by+y_0)+S\cosh(4by+y_0)	\eqno(4.7)$$
with $M={1\over{2b^2}}\sqrt{m^2-4b^2n}$, $m_0={m\over{2b^2}}$,
 $s_0={{3s}\over{4b^2}}$, 
$S={3\over{8b^2}}{{sm+2b^2h}\over\sqrt{m^2-4b^2n}}$,
where $m$, $n$, $s$ and $h$ are the constants used in 
the field equations of Ref. [1] and, finally, $x_0$, $y_0$, $N_1$
and $N_2$ are constants of integration 
\footnote{$^4$}{\sevenrm Note that the functional form of 
{\it W(y)} given by eq. (4.5) 
differs from 
the original expression of {\it W(y)} in Ref. [1]. 
After checking the 
functional form of {\it W(y)} given by eq. (21) of Ref. [1]
we have found that it does not satisfy eq. (18) of Ref. 
[1] unless the 
second term of the right-hand side of eq. (21) of Ref. [1] is divided by 
the constant $C_1$ used there. Since the function 
{\it H(y)} depends on 
the explicit form of {\it W(y)} (see eq. (19b) of Ref. [1]),
to have the correct solution given by eq. (4.7) for class I, one 
has to alter the function {\it H(y)} correspondingly. Furthermore, 
in the last term of the expression for {\it G(x)} for 
class I in 
Ref. [1] instead of the factor $(4bx+C_4)$ there should be 
$(-4bx+C_4)$ in order to have the correct solution of the basic field 
equations.}.

For class II:
$$V(x)={m\over{2b^2}}=constant			\eqno(4.8)$$
$$W(y)=L_1e^{-4by}+{m\over{2b^2}}		\eqno(4.9)$$
$$G(x)={h\over{16b^2}}-L_3\cos (4bx)-L_4\sin(4bx)\eqno(4.10)$$
$$H(y)={h\over{16b^2}}+L_2e^{-4by}		\eqno(4.11)$$
where $L_1,L_2,L_3$ and $L_4$ are arbitrary constants.

Using eqs. (4.2)-(4.7) the norm of the Killing field, $v_0$, and the 
twist potential, $\o_0$, for class I can be shown to be
$$v_0=-{1\over M}\Biggl[S+
{N_1\cos (4bx+x_0)-N_2\sinh (4by+y_0)
\over{\sin (4bx+x_0)-\cosh (4by+y_0)}}\Biggr],
\eqno(4.12)$$
$$\o_0={1\over M}{{N_2\cos (4bx+x_0)+
N_1\sinh (4by+y_0)}\over{\sin (4bx+x_0)-\cosh (4by+y_0)}}.
\eqno(4.13)$$

The same quantities for class II are
$$v_0=-{1\over L_1}\bigl(L_3\cos (4bx)+L_4\sin (4bx)\bigr)
e^{4by}-{L_2\over L_1},			\eqno(4.14)$$
$$\o_0={1\over L_1}e^{4by}\bigl(-L_3\sin (4bx)+L_4\cos (4bx)\bigr).
\eqno(4.15)$$
Here the twist potentials have been determined from the relevant form of eq. (16.23) 
of Ref. [13].

The families of perfect-fluid configurations obtained from the 
Senovilla-metrics
are given as
$${ds_\tau}^2=v_\tau(dt+A_\tau d\phi)^2-{1\over v_\tau}
\Bigl[{{G-H}\over G}dx^2+{{G-H}\over H}dy^2 +c^2GHd\phi^2\Bigr]	,
\eqno(4.16)$$
where G and H are the original functions given by eqs.
(4.6)-(4.7) and eqs. (4.10)-(4.11) for class I and class II,
respectively. Moreover, the actual forms of $v_\tau$ and $\o_\tau$ 
can be calculated by using eqs. (3.1) and (3.2)
with $v_0$ and $\o_0$ determined by eqs. (4.12) and (4.13) for class I
and by eqs. (4.14) and (4.15) for
class II. Finally, the function $A_\tau=A_\tau (x,y)$ is
the solution of the system
$${{\d A_\tau}\over{\d x}}=cH{v_\tau}^{-2}{{\d\o_\tau}\over{\d y}},
\eqno(4.17)$$
$${{\d A_\tau}\over{\d y}}=-cG{v_\tau}^{-2}{{\d\o_\tau}\over{\d x}}.
\eqno(4.18)$$

Using expression (3.1) for $v_{\tau}$, it turns out that the general solution of eqs.
(4.17) and (4.18) can be written as
$$A_\tau={{HW_\tau -GV_\tau}\over{H-G}}		\eqno(4.19)$$
where $V_\tau$ and $W_\tau$ are the transformed versions of the functions $V(x)$ and 
$W(y)$. It is also true, that $v_\tau$ can be expressed as 
$v_\tau=(H-G)/(V_\tau-W_\tau)$. The functions $V_\tau$ and $W_\tau$ are
$$V_\tau (x)=f(\tau)+C_{1\tau} \sin(4bx+x_0)+C_{2\tau}\cos(4bx+x_0)\eqno(4.20)$$
$$W_\tau (y)=f(\tau)+C_{3\tau}\sinh(4by+y_0)+C_{4\tau}\cosh(4by+y_0) \eqno(4.21)$$
where $f(\tau)$ is an arbitrary function of $\tau$ with the property 
$f(0)=m_0$ while the factors $C_{1\tau}$, $C_{2\tau}$, $C_{3\tau}$ and 
$C_{4\tau}$ are given by the following expressions.

For class I:
$$C_{1\tau}={1\over M}\bigl[\sin^2\tau (S^2-N_1^2-N_2^2)+M^2\cos^2\tau 
\bigr],\eqno(4.22)$$
$$C_{2\tau}={{2\sin\tau}\over M}\bigl[\sin\tau N_1S-\cos\tau N_2M\bigr],\eqno(4.23)$$
$$C_{3\tau}={{2\sin\tau}\over M}\bigl[\sin\tau N_2S+\cos\tau N_1M\bigr],\eqno(4.24)$$
$$C_{4\tau}={1\over M}\bigl[\sin^2\tau (S^2+N_1^2+N_2^2)+M^2\cos^2\tau 
\bigr].	\eqno(4.25)$$

For class II:
$$C_{1\tau}=-{{2\sin\tau}\over L_1}(L_2L_4\sin\tau+L_1L_3\cos 
\tau)	,			\eqno(4.26)$$
$$C_{2\tau}=-{{2\sin\tau}\over 
L_1}(L_2L_3\sin\tau-L_1L_4\cos\tau),\eqno(4.27)$$
$$C_{3\tau}={{2(L_1^2+L_3^2)}\over L_1}\sin^2\tau-L_1,
\eqno(4.28)$$
$$C_{4\tau}=-{{2(L_2^2+L_3^2)}\over L_1}\sin^2\tau+L_1	.
\eqno(4.29)$$

It is straightforward to check that for both classes I and II 
the identity $C_{4\tau}^2=C_{1\tau}^2+C_{2\tau}^2+C_{3\tau}^2$ 
holds. Using this relationship, it can be seen that the transformed 
solutions $V_\tau$, $W_\tau$, $G$ and $H$ also satisfy the 
original field equations (16) and (17) of Ref. [1] 
\footnote{$^5$}{\sevenrm See 
also the general form of the functions $V$, $W$, $G$ and $H$ given in 
appendix 2.}. Therefore, we 
can state that the transformed metrics represent Petrov type D 
stationary axisymmetric rigidly-rotating perfect-fluid 
configurations, i.e. the solutions obtained by the application
of the transformation are, actually, not new.
In particular, it can be shown that
the metrics obtained from 
class I include (for special settings of the parameters) {\it both} of 
the original classes (${dV\over{dx}}\not =0$ and ${dV\over{dx}}=0$) of the Senovilla metrics, while the metrics 
obtained from class II comprise a subclass of this family.
Note that although the transformation does not yield new solutions from 
the Senovilla metrics, it is not a gauge transformation. This 
follows from the fact that it maps both class I and class II 
onto a strictly larger subclass of the entire family 
of solutions.

\bigskip
\parindent 0pt
{\bf 5. Conclusions}
\medskip
\parindent 20pt

Stationary perfect-fluid configurations 
with 4-velocity parallel to the stationary Killing field were 
considered for which the norm and the twist potential of the Killing
field are functionally independent. It was shown that by the use of 
`effective' $SL(2,$\re$)$ transformations new perfect-fluid solutions can be
obtained from known ones whenever the equation of state is $\rho+3p=0$.
By applying the relevant method to known geometries 
with this equation of state
we obtained stationary perfect-fluid solutions of Einstein's 
equations which contain in a compact form  all the configurations we used as input 
for our procedure. Although these solutions are not gauge related to the ones
we started with, there are no new perfect-fluid configurations among them, since 
all special cases were given previously. 

Note, finally, that the perfect-fluid solutions to 
which our 
procedure has been applied here possess at least two commuting Killing fields.
However, we would like to emphasize that our method postulates merely the 
existence of a single timelike Killing field.

\bigskip\parindent 0pt
{\bf Appendix 1}
\medskip\parindent 20pt

Here it is shown that the two families of 
metrics given by formulas (27) and (42) of Ref. [3] 
are, in 
fact, subclasses of class II of Senovilla metrics.

\bigskip
{\it The case of a Mars-Senovilla metric with two timelike Killing 
fields}

A stationary rigid perfect-fluid solution with 
equation of state $\r+3p=0$ was
found by Mars and Senovilla [3].
This spacetime is also of Petrov type D and its line element is 
$$ds^2={{{\a\b}\over{a^2y^2}}dt^2-{x^2\over{\a\b}}\bigl(aydT+
{\a\over{ay}}dt\bigr)^2+{{a^2y^2dx^2}\over{\delta^2-
{{a^2x^2}\over\b^2}(x^2-2\b^2)}}+dy^2},		\eqno(A1.1)$$
where $\a>0$, $\b>0$, $\delta$ and $a$ are arbitrary constants.
This solution possesses two commuting Killing vector fields,
$\bigl({\d\over{\d t}}\bigr)^a$ and 
$\bigl({\d\over{\d T}}\bigr)^a$,
associated with the coordinates $t$ and $T$, respectively.
It is an interesting feature of this solution that both of the commuting 
Killing fields are timelike. The 4-velocity of the fluid is
parallel to the Killing field  $\bigl({\d\over{\d t}}\bigr)^a$.

To show that this metric is  isometric to
class II of Senovilla metrics change the variables as
$$dt\rightarrow\sqrt{\alpha\over\beta}dt \ ,\ 
dx\rightarrow\sqrt{G}dx\ ,\ 
dy\rightarrow\sqrt{H}dy\ ,\ 
d\phi\rightarrow dT	\eqno(A1.2)$$
and set the functions in the Senovilla metrics to
$$G(x)=Q^2(x),\ \ V(x)=0,\ \ H(y)=\beta^2=constant,\ \ W(y)=-e^{-2ay},
\eqno(A1.3)$$
where $\alpha$ and $\beta$ are constants of the Mars-Senovilla 
metric. Substituting the above expression for $G(x)$ in eq. 
(20a) of Ref. [1] and changing the constants of Ref. [1] to 
$$h=4a^2,\ \  b={1\over 2}{a\over \beta},\ \ c={1\over\sqrt{\alpha\beta}}	
\eqno(A1.4)$$
we get the corresponding equation of Ref. [3] for $Q(x)$ as given 
at the bottom of page 3063 in Ref. [3].  
From this point, the same way as in Ref. [3], one obtains 
the metric in the form given by eq. (42) of Ref. [3].

Finally, note that the ``axial" Killing vector of the Senovilla 
metrics, with the settings (A1.2)-(A1.4), becomes, actually, a 
timelike Killing vector.

\bigskip
{\it The case of a stationary axisymmetric Mars-Senovilla metric}

Another stationary axisymmetric Petrov type D perfect-fluid
solution with equation of state $\r+3p=0$ was also
found by Mars and Senovilla (see case I of Ref. [3]). Originally,
this
solution was thought of and described as a differentially rotating 
perfect fluid. However,
the authors informed us, that, although the line element is correct (see 
eq. (27) of Ref. [3]), this stationary perfect-fluid solution is,
in fact, rigidly rotating 
with 4-velocity  parallel to the stationary Killing field.

The line element of this spacetime is 
$$ds^2={-\b^2 \dot H^2dt^2+X^2\biggl({1\over H}d\phi+\b\o 
Hdt\biggr)^2+{{dX^2}\over{H^2(2\epsilon a^2 X^2-\o^2X^4+L_0)}}+
{{dy^2}\over{H^2}}},			\eqno(A1.5)$$
where $\b$, $\o $, $a$ and $L_0$ are arbitrary constants and
$\epsilon$ is a sign [3]. Here $H$ is a function of the variable $y$ 
satisfying
$$ {\ddot H}={\epsilon a^2H},		\eqno(A1.6)$$
where the dot denotes derivative with respect to $y$.

To show that this metric can be obtained from class II of
Senovilla metrics,
first, change the variables as
$$dt\rightarrow{{dt}\over\sqrt{\beta}} \ ,\ 
dx\rightarrow\sqrt{G}dx\ ,\ 
dy\rightarrow\sqrt{H_{_S}}dy\eqno(A1.7)$$
in the Senovilla metrics where $H_{_S}(y)$ is the function given by eq. (4.11). 
Furthermore, set
$$G(x)=\omega^2L^2(x),\ \ V(x)=0,\ \ 
H_{_{S}}(y)={{\dot H^2(y)}\over{H^2(y)}},\ \ W(y)=-H^{-2}(y),
\eqno(A1.8)$$
where $\beta$ and $\omega$ are constants of the Mars-Senovilla 
metric. Then,
setting the constants of the Senovilla metrics as
$$ h=4\epsilon a^2,\ \ b={1\over 2},\ \ c={1\over\omega},	\eqno(A1.9)$$
one can see that the 
functions $G(x)$, $V(x)$, $H_{_{S}}(y)$ and $W(y)$ satisfy eqs. (18) 
and (20a-b) of Ref. [1] whenever the function $L(x)$ satisfies eq. 
(25) of Ref. [3] and 
$H(y)$ satisfies the equation given bellow eq. (27) 
of Ref. [3].

\bigskip\parindent 0pt
{\bf Appendix 2}
\medskip\parindent 20pt

Here we are going to reflect briefly on a method yielding solutions of the 
field equations (16) and (17) of Ref. [1]. This way one gets the general 
form of all stationary axisymmetric Petrov type D rigidly rotating 
perfect-fluid metrics with an equation of state $\rho+3p=constant$ 
possessing a simpler functional form than the original solutions given in 
Ref. [1].

The metric is given by eq. (4.1) and the basic field equations for the 
functions $V=V(x)$, $W=W(y)$, $G=G(x)$ and $H=H(y)$ 
(see eqs. (16) and (17) of Ref. [1]) are
$$V'^2+W'^2=2(V-W)[8b^2(V-W)+V''], \eqno(A2.1)
$$
$$ 12a(V-W)=16b^2(G-H)+V''{{G-H}\over 
{V-W}}-{{V'G'+W'H'}\over {V-W}}+G'',\eqno(A2.2) $$
where the prime denotes derivation with respect to the argument.
The general solutions of these equations 
can be obtained as follows.

Taking the derivative of eq. (A2.1) with respect to the 
variable $x$ and using the fact that for a
non-singular metric $V-W\not =0$ we get
$$ (V'''+16b^2V')=0. \eqno(A2.3)$$

The general solution of this equation is
$$ V(x)=V_0+C_1\sin(4bx)+C_2\cos(4bx), \eqno(A2.4)$$
where $V_0$, $C_1$ and $C_2$ are constants.

By a similar procedure, calculating the derivative of eq. 
(A2.1) with respect to $y$, we get

$$ W'(W'''-16b^2W')=0 \eqno(A2.5)$$
which holds either if 
$$ W'=0 \eqno(A2.6) $$
or 
$$ W'''-16b^2W'=0. \eqno(A2.7) $$

The general solution of eq. (A2.7) for the case when $W' 
\not = 0$ is
$$ W(y)=W_0+C_3\sinh(4by)+C_4\cosh(4by). \eqno(A2.8) $$

The substitution of expressions (A2.4) and (A2.8) into 
eq.(A2.1) yields the following restrictions on the 
constants
$$ V_0=W_0=:C_0 \ \ {\rm and}\ \  
C_4^2=C_1^2+C_2^2+C_3^2 .\eqno(A2.9) $$

Substituting  $W'=0$ and eq. (A2.4) into eq. (A2.1) 
we have
$$ W(y)=C_0\pm \sqrt{C_1^2+C_2^2} =const\ .
\eqno(A2.10) $$

To get the solutions to eq. (A2.2), first note that by 
using
$$ V''+16b^2V=16b^2C_0 \eqno(A2.11)$$
eq. (A2.2) can be recast into the form
$$ 12a(V-W)=16b^2(G-H)(C_0-W)-V'G'-W'H'+G''(V-W) .
\eqno(A2.12) $$

Taking the derivative of eq. (A2.12) with respect 
to $x$, using  (A2.11) and the fact
that $V-W\not =0$ we arrive at
$$ G'''+16b^2G'=24aV' .\eqno(A2.13) $$

This equation has the general solution 
$$
G(x)=G_0+K_1\sin(4bx)+K_2\cos(4bx)-{{3a}\over{4b^2}}xV',
\eqno(A2.14) $$
where $G_0$, $K_1$ and $K_2$ are constants of integration.

In order to determine the function $H(y)$ we take the derivative 
of eq. (A2.12) with respect to $y$ and use
$$ W''-16b^2W=16b^2C_0. \eqno(A2.15)$$

This yields
$$ 24a(V-W)W'=W'(G''+16b^2G+H''-16b^2H). \eqno(A2.16)$$

First, consider the case when $W'\not =0$. Then from eq. (A2.16)
it follows that
$$ H'''-16b^2H'=-24aW' \eqno(A2.17)$$
the general solution of which is
$$
H(y)=H_0+K_3\sinh(4by)+K_4\cosh(4by)-{{3a}\over{4b^2}}yW',
\eqno(A2.18) $$
where $H_0$, $K_3$ and $K_4$ are constants of integration.

By the substitution of expressions (A2.14) and (A2.18) for $G(x)$ and 
$H(y)$ into eq. (A2.15) one can see that the constants of integration 
have to satisfy the following conditions 
$$H_0=G_0=:K_0\ \ {\rm and}\ \ 
K_4C_4=K_1C_1+K_2C_2+K_3C_3. \eqno(A2.19)$$

Finally, whenever $W'=0$ eq. (A2.16) is satisfied identically. Then 
the substitution of
$W=C_0\pm\sqrt{C_1^2+C_2^2}$\  into eq. (A2.12) yields
$$ H(y)=K_0\pm {{C_1K_1+C_2K_2} \over 
\sqrt{C_1^2+C_2^2}}=constant\ . \eqno(A2.20) $$

Our solutions can be related to the original Senovilla ones by setting
the constants $m$, $n$, $s$ and $h$,
introduced in the field equations (18)-(20)
of Ref. [1], in terms of our constants of integration as follows:
$$\eqalign{m&=2b^2C_0,\ s={4\over 3}b^2K_0-2aC_0,\ 
n=b^2(C_0^2-C_1^2-C_2^2),\cr h&=a(C_0^2-C_1^2-C_2^2)-{4\over 3}b^2(
C_0K_0-C_1K_1-C_2K_2).}\eqno(A2.21)$$

Finally, note that the class II Senovilla metrics can be recovered from our solutions by 
choosing the constants $C_1=C_2=0$.

\bigskip\parindent 0pt
{\bf Acknowledgments}
\medskip\parindent 20pt

One of us (I.R.) wishes to thank J. M. M. Senovilla for
useful discussions and also for drawing into our 
attention a number of solutions which could be used in our 
formalism. Thanks are also due to \'A. Sebesty\'en for helpful discussions.
A great number of calculations has been carried out using 
the algebraic computer programme MAPLE V
and also by making use of GRTensor II [18].
This research was supported in parts by the OTKA grants F14196
and T016246.

\vfill\eject

\medskip\parindent 0pt
{\bf References}
\medskip

\item{[1]} J. M. M. Senovilla: Phys. Lett.  {\bf A 123} number 5,
211-214 (1987)

\item {[2]} J.M. M. Senovilla: Class. Quant. Grav. {\bf 4},
L115-L119 (1987)

\item{[3]} M. Mars and J. M. M. Senovilla: Class. Quantum Grav.
{\bf 11} 3049-3068 (1994)
 
\item{[4]} S.Bonanos and E. Kyriakopoulos: Class. Quantum Grav.
{\bf 11} L23-L28 (1994)

\item{[5]} D.Kramer: Class. Quantum Grav. {\bf 1} L3-L7 (1984) 

\item{[6]} D.Kramer: Class. Quantum Grav. {\bf 2} L135-L139 (1985) 

\item{[7]} E. Kyriakopoulos: Gen. Rel. Grav. {\bf 20}, 427-436, 
(1988)

\item{[8]} F. J. Chinea and  L. M. Gonzales-Romero: Class. Quant. 
Grav. 1271-1301 (1992)

\item{[9]} R. Geroch: J. Math. Phys. {\bf 12}, 918-924 (1971)

\item{[10]} I. R\'acz: ESI Preprint Series No. 168/94 (1994)
 
\item{[11]} I. R\'acz: in {\it Inhomogeneous Cosmological Models}, 
eds. A. Molina and J. M. M. Senovilla (World Scientific, Singapore, 
1995)

\item {[12]} D. Kramer, G. Neugebauer and H. Stephani: Fortschr. Physik, 
{\bf 20} 1-75 (1972)

\item{[13]} D. Kramer, H. Stephani, M. MacCallum and E. Herlt: 
{\it Exact solutions of Einstein's field equations} (Cambridge;
Cambridge University Press, 1980)

\item{[14]} D.Kramer, G. Neugebauer: Annalen Physik {\bf 27}
 129-135 (1971)

\item{[15]} R. Geroch and L. Lindblom: Ann. Phys. {\bf 207},
394-416 (1991)

\item{[16]} C. M. Cosgrove: J. Phys. A {\bf 11}, 2389-2404 (1978)

\item{[17]} E. D. Fackerell and R. P Kerr: Gen. Rel. Grav. {\bf 
23}, 861-876 (1991)

\item{[18]} P. Musgrave, D. Pollney and K. Lake: {\it GRTensor II},
(Queens University, Kingston, Ontario, Canada 1994)
 
\vfill\eject

\end